\documentclass{article}  
\usepackage{breckenridge}
\frompage{000} \topage{000}                                              
\def\rightmark{B Physics at STAR}
\def\leftmark{Jens S\"oren Lange}

\title{B Physics at STAR} 
\authors{
{Jens S\"oren Lange$^1$, for the STAR Collaboration}\\[2.812mm]
{\normalsize
\hspace*{-8pt}$^1$ University of Frankfurt, Institut f\"ur Kernphysik, \\
August-Euler-Stra\ss{}e 6, 60486 Frankfurt/Main, Germany\\[0.2ex] 
}}
 
 
\begin{document}
 
\maketitle
\setcounter{page}{1}

\section{$b$ quarks in $A$+$A$ collisions}

In relativistic nucleus-nucleus collisions, 
heavy quarks ($c$,$b$) are produced in the early stage \cite{early}, 
at time scales $t$$\simeq$0.2-0.5~fm/c, in comparison
to typical pion freeze-out time scales of $t$$\simeq$10~fm/c.
In particular, $b$ quarks are interesting for two reasons:\\
\noindent {\it (1)} The mass of the $c$ quark $m$($c$)=1.40~GeV 
is about the same order of magnitude as $\Lambda_{QCD}$$\simeq$1~GeV,
but $m$($b$)=4.75~GeV is significantly larger.
Thus, for the $b$ quarks, pQCD based 
cross section predictions 
are more reliable than for $c$ quarks.\\
\noindent {\it (2)} For $Au$+$Au$ collisions at RHIC, the primordial temperature 
might be high enough to create $c$$\overline{c}$ pairs thermally, 
but not $b$$\overline{b}$ pairs. For $t$=0.6~fm/c, a temperature 
of $T$$\simeq$0.3~GeV can be estimated \cite{kolb}.
At such a high temperature, QCD predicts that 
gluons, which are massless at $T$=0,
acquire an effective mass, which turns out to 
be in the same order of magnitude as the temperature itself, 
i.e.\ $m_{gluon}$$\simeq$0.3~GeV \cite{vogt1}.
Thus the production process $g$$g$$\rightarrow$$c$$\overline{c}$
might be enhanced thermally.
However, for $b$$\overline{b}$ production, 
the Boltzmann suppression factor is
\mbox{$<$$b$$\overline{b}$$>$/$<$$c$$\overline{c}$$>$=$exp$$[$$-$2$m$($b$)/2$m$($c$)$]$}\\
$\simeq$1/23, thus any thermal $b$$\overline{b}$ enhancement is unlikely.
Thus, $b$ quarks can be regarded as a pure probe 
of the early stage of the collision.
However, $b$$\overline{b}$ events are rare.
At RHIC, $b$$\overline{b}$ measurements are now feasible for first time,
due to three reasons: 
{\it (1)} The $b$$\overline{b}$ production cross section at $\sqrt{s}$=200~GeV (RHIC)
is about a factor of $\sim$10$^3$ higher than at $\sqrt{s}$=18~GeV (SPS). 
{\it (2)} High collider luminosity provides a recorded data sample 
of 10$^6$-10$^7$ events. 
{\it (3)} Recent technological and price development of fast CPUs created
the basis for the possibility of event reconstruction in realtime
for application as high level trigger on rare events.

\noindent This paper describes the first attempt to observe $b$$\overline{b}$
production with the STAR experiment at RHIC.
The main STAR detector is a large scale ($R$=2~m, $L$=4~m) TPC (Time Projection
Chamber) \cite{star} with 5$\times$10$^7$ position pixels, each measuring the particle 
ionization dE/dx using a 10-bit ADC.
For triggering on rare $b$$\overline{b}$ events,
a Level-3 trigger CPU farm was used, described in Section~\ref{cl3}.

\noindent The $b$$\overline{b}$-bound state $\Upsilon$ 
is specifically interesting, as the formation and re-breakup rates 
are sensitive to color screening and thus the initial gluon density.
STAR is particularly suited for the search for $\Upsilon$ production
due to its large midrapidity coverage. Because of its high mass, 
the $\Upsilon$ is produced almost at rest, disfavouring 
gluon-boosted forward/backward emission of the decay products.


\section{The case of the $\Upsilon$}\label{upsilon}
\label{cupsilon}

In the following, the term $\Upsilon$ will be used to represent the
three $S$ states $\Upsilon$(1$S$), $\Upsilon$(2$S$) and $\Upsilon$(3$S$).
$\Upsilon$ properties are listed in Tab.~1. 
The production ratio is independent from $\sqrt{s}$
in leading order QCD, thus ratios measured at $\sqrt{s}$=1.8~TeV
are applicable to RHIC energies ($\sqrt{s}$=200~GeV).\\
\begin{minipage}[b]{6.0cm}{
\begin{tabular}{|l|r|r|r|}
\hline
& $m$/GeV & $R_{prod}$ & BR($e^+$$e^-$)\\
\hline
$\Upsilon$(1$S$) & 9.460 & 72\% & 2.38\% \\
$\Upsilon$(2$S$) & 10.023 & 18\% & 1.18\% \\
$\Upsilon$(3$S$) & 10.355 & 10\% & 1.81\% \\
\hline
\end{tabular}
}
\end{minipage}
\hfill
\begin{minipage}[b]{6.0cm}{
{\bf Table 1.} Mass $m$ \cite{pdg}, 
production ratio for midrapidity $R_{prod}$ \cite{cdf} and 
branching fraction \cite{pdg} into $e^+$$e^-$,
for the lowest 
three $\Upsilon$ states.
}
\end{minipage}

\noindent The $\Upsilon(4s)$ and $\Upsilon(5s)$ shall not be considered here,
as they are above the $B$$\overline{B}$ 
and $B_s$$\overline{B_s}$ thresholds, respectively.
All $\Upsilon$ singlet $^{1}S$, $^{1}P$ states are still unobserved,
although they are expected to exist. So far, only $\Upsilon$ triplet states 
have been observed, because most data were recorded from 
$e^+$$e^-$ annihilation ($J^{PC}$($\gamma^*$)=1$^{--}$).
Recently the discovery of $\Upsilon$(1$D$) was reported \cite{cleo},
the only long-lived $L$=2 meson ever observed.
The $\Upsilon$ decays into $e^+$$e^-$ with a branching fraction
of $\simeq$2\% (Tab.~1).
Experimentally, this is the easiest exclusive final state to measure,
whereas the decay into a {\it ggg} final state (3 jets) 
with the largest branching fraction of 
77.4\% is experimentally unobservable.
At $\sqrt{s}$=200~GeV, 
$\simeq$90\% of the $\Upsilon$ are produced by $g$$g$$\rightarrow$$b$$\overline{b}$, 
$\simeq$10\% by $q$$\overline{q}$$\rightarrow$$b$$\overline{b}$ ($q$ denoting $u$, $d$, $s$ or $c$).
As the total cross section for 
$b$$\overline{b}$ production in $p$+$p$ is
$\sigma$($b$$\overline{b}$)=1.5~$\mu$b,
one expects about one $b$$\overline{b}$ 
pair in $\simeq$52 $Au$+$Au$ central events.
With a production cross section of $\sigma$($\Upsilon$)=8.4~nb, 
it follows, that 1/178 of $b$$\overline{b}$ pairs form a resonance\footnote{
As a comparison, 1/116 $c$$\overline{c}$ pairs form a J/$\psi$.}.
A pure statistical production model, which works well for lighter 
particles, does not seem to be applicable \cite{marek}.
For a rate estimate,
the baseline prediction for $\sigma_{pp}$($\Upsilon$)
is taken from \cite{vogt2}. The cross section has a typical threshold 
behaviour $\sigma$=$\sigma_o$(1$-$$m_{\Upsilon}$/$\sqrt{s}$)$^n$,
rising steeply in the RHIC energy region.
Feed-down contributions are included in the calculation,
e.g.\ only 52\% of the $\Upsilon$(1$S$) are produced directly.
Feed-down contributions are 
10\% from the $\Upsilon$(2$S$),
2\% from the $\Upsilon$(3$S$),
26\% from the $\chi_b$(1$P$),
and 10\% from the $\chi_b$(2$P$).
With an absorption correction $\alpha$, 
the fraction of the geometrical cross section $f_{geo}$ (10\% for central collisions)
and the fraction of hard processes $f_{hard}$,
the number of expected $\Upsilon$ per one central $Au$+$Au$ collision
at $\sqrt{s}$=200~GeV can be estimated as

\vspace*{-0.3cm}
\begin{eqnarray}
N_{\Upsilon}^{Au+Au} & = &
BR(e^+e^-)\cdot\frac{d\sigma_{pp}}{dy}|_{y=0} \cdot
\frac{\Delta y 
\cdot \varepsilon_{rec}
\cdot (AA)^{\alpha} \cdot f_{hard}}
{\sigma_{geo} \cdot f_{geo}} \\
 & = & 100 pb \cdot \frac{2 \cdot 14.5\% \cdot (197\cdot 197)^{0.95} \cdot 0.4}{7.2 b \cdot 0.1}
\simeq 3.7\cdot 10^{-7} . \nonumber
\end{eqnarray}


\section{Lepton Identification}

\noindent Electrons were identified by their specific ionization dE/dx in the TPC,
which saturates at dE/dx$\simeq$17-18~keV/cm for $p$$\geq$1~GeV/c.
Due to the mass dependance in the relativistic Bethe-Bloch formula, 
charged pions show increasing dE/dx$\simeq$11-14~keV/cm for $p$=1-10~GeV/c.
As shown in Fig.~\ref{fig1}, for this analysis electron candidates 
were identified by a cut on dE/dx$\geq$15~keV/cm.

\quad\\
\begin{minipage}[b]{6.1cm}{
\leavevmode
\epsfysize=4.0cm
\epsfxsize=6.3cm
\epsfbox{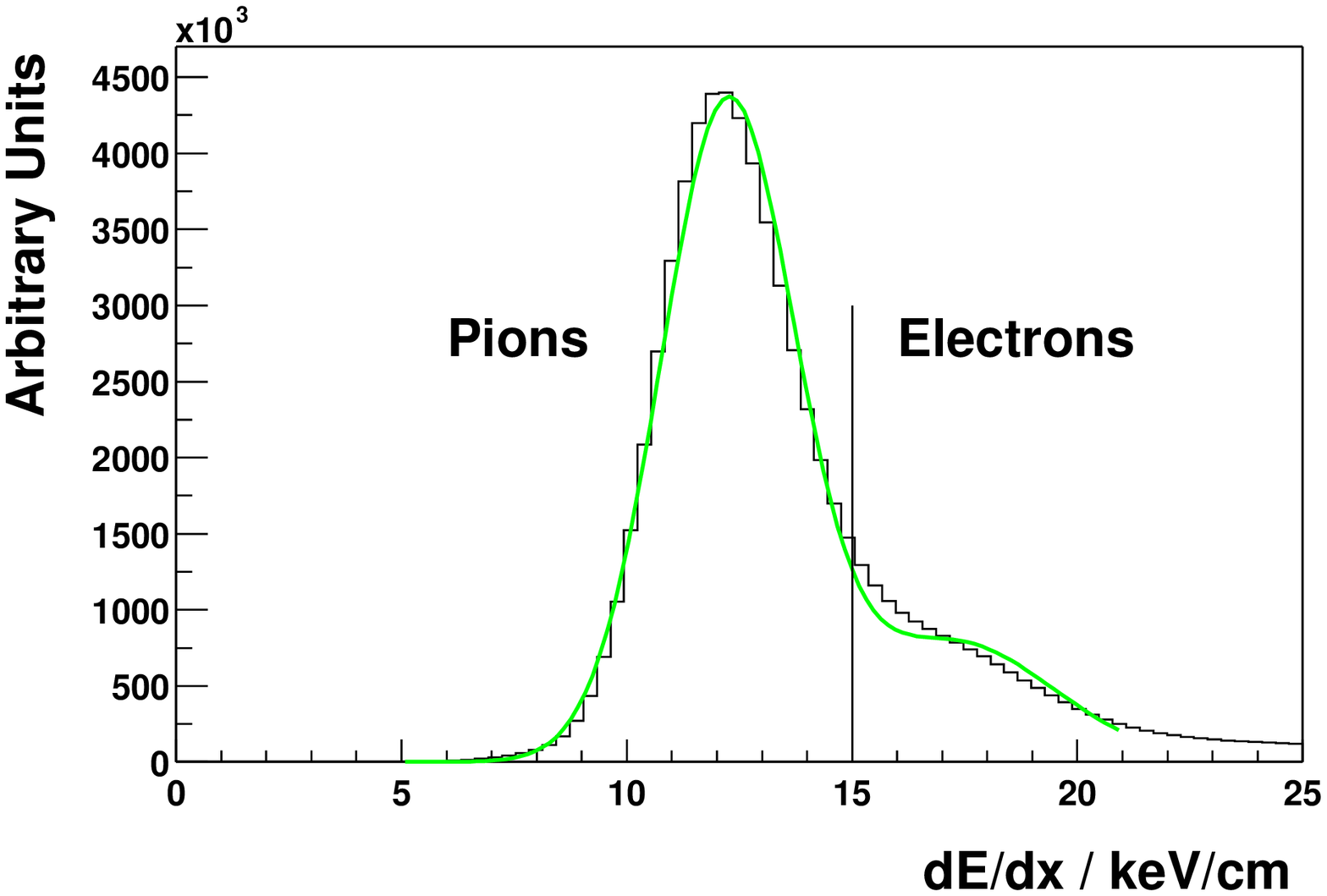}
\label{fig1}}
\end{minipage}
\leavevmode
\epsfysize=4.0cm
\epsfxsize=6.3cm
\epsfbox{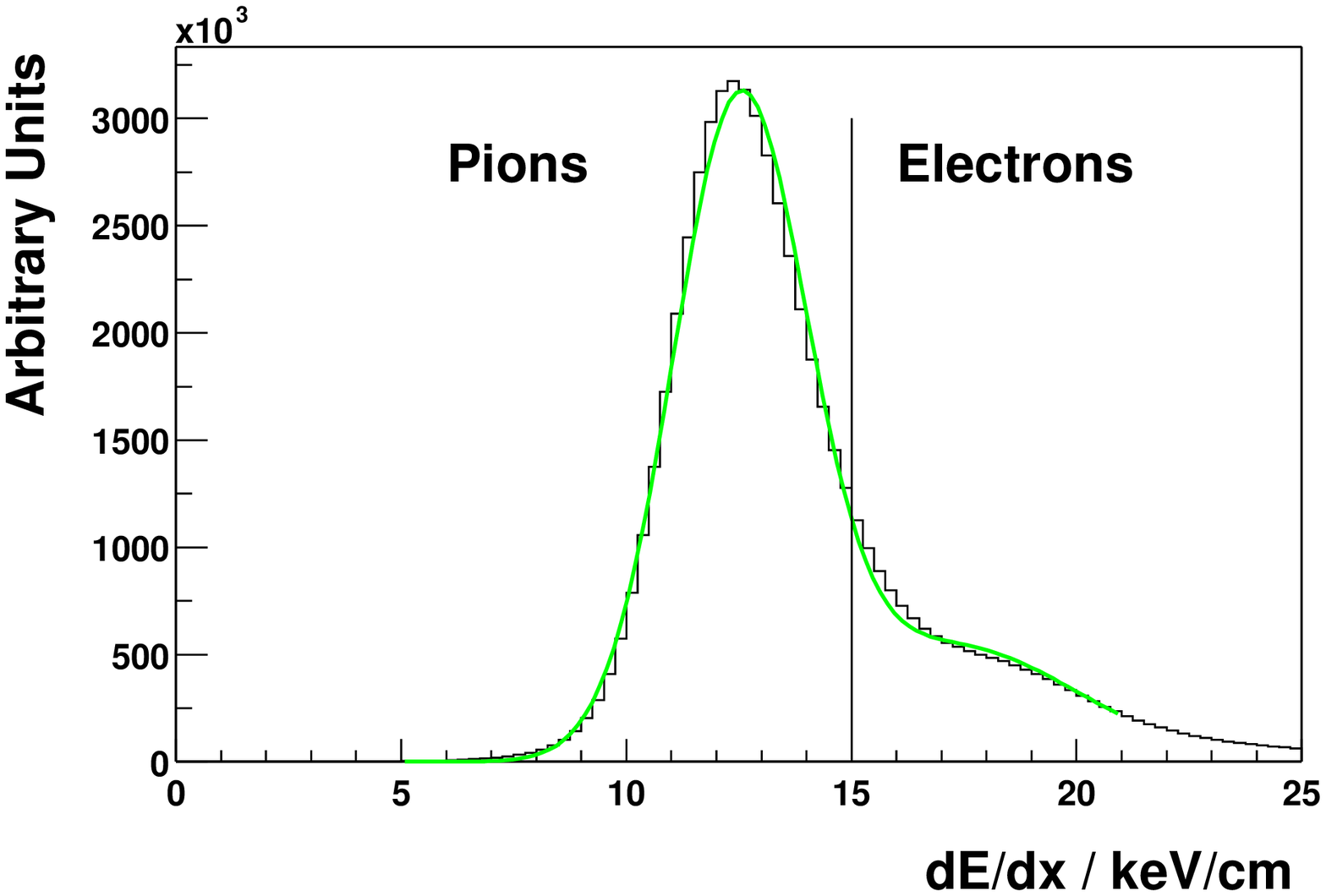}
\begin{minipage}[b]{6.1cm}{
\quad\\ \quad\\ \quad\\ \quad\\ \quad\\ \quad\\ }
\end{minipage}
\vspace*{-3em}

\noindent {\bf Fig.~1.}~TPC ionization dE/dx for Level-3 trigger {\it (left)}
and offline {\it (right)} for $p_T$$\geq$3~GeV. 
For electron identification, a cut dE/dx$\geq$15~keV/cm was confirmed.

\vspace*{1em}

\noindent The observed $e$/$\pi$ ratio for events with two $p$$\geq$3~GeV/c tracks 
is $\simeq$1/3.5, determined from a two-gaussian fit (line in Fig.~1). 
The pion contamination in the electron sample is 17\%. Imperfection of the fit in the region
dE/dx$\simeq$14-16~keV/cm is resulting from muon contamination, which is not
taken into account for this paper\footnote{For dilepton invariant mass in the high 
mass region, mass differences between electrons and muons are negligible, 
and thus, for the muon contamination, an identical invariant mass slope is expected.}.
Additional contamination from other particles (e.g.\ deuterons) is included
in the systematic error. The electron identification was additionally checked 
with a partial electromagnetic calorimeter \cite{emc} (400 towers, 
$\Delta$$\varphi$=60$^o$, 0$\leq$$\eta$$\leq$1).
For dE/dx-selected electron candidates $p$/$E$=1.0$\pm$0.2 was achieved.


\section{The Level-3 Trigger}
\label{cl3}

The STAR Level-3 trigger performs full-event reconstruction
in realtime, utilizing both fast CPUs and a fast network.
432 Intel i960 CPUs perform TPC cluster finding,
48 ALPHA DS-10 CPUs perform TPC track finding,
and 3 Intel Pentium CPUs apply global algorithms,
such as invariant mass calculations.
For the 10\% most central $Au$+$Au$ events ($\simeq$130.000 clusters, $\simeq$4500 tracks),
a sustained Level-3 processing rate of $R$$\leq$40~Hz was achieved.
For a solenoidal magnetic field of $B$=0.25~T,
the $p_T$ resolution for (electron) tracks with $p_T$=5 GeV/c 
was determined offline to be $\Delta$$p_T$/$p_T$$\simeq$9\% and $\simeq$17\%
(with and without vertex constraint, respectively), 
and, on the Level-3 trigger, $\Delta$$p_T$/$p_T$$\simeq$10\% and $\simeq$26\% \cite{l3}.
For $B$=0.5~T, all resolutions are to be multiplied by a factor 1/3.
The dE/dx resolution is offline $\simeq$8\%, 
and on the Level-3 trigger $\simeq$11\%.
Further details are described elsewhere \cite{l3}.
For the $\Upsilon$, the Level-3 trigger algorithm performed
a trigger YES/NO decision, based upon an invariant mass cut
for $p_T$$\geq$3~GeV/c and dE/dx$\geq$15~keV/cm selected 
electron candidates. The $\Upsilon$ enhancement factor 
varied in the range 1.8-4.1, based on momentary accelerator
luminosity conditions.


\section{Results}
\label{cresults}

\begin{figure}[htb]
\insertplot{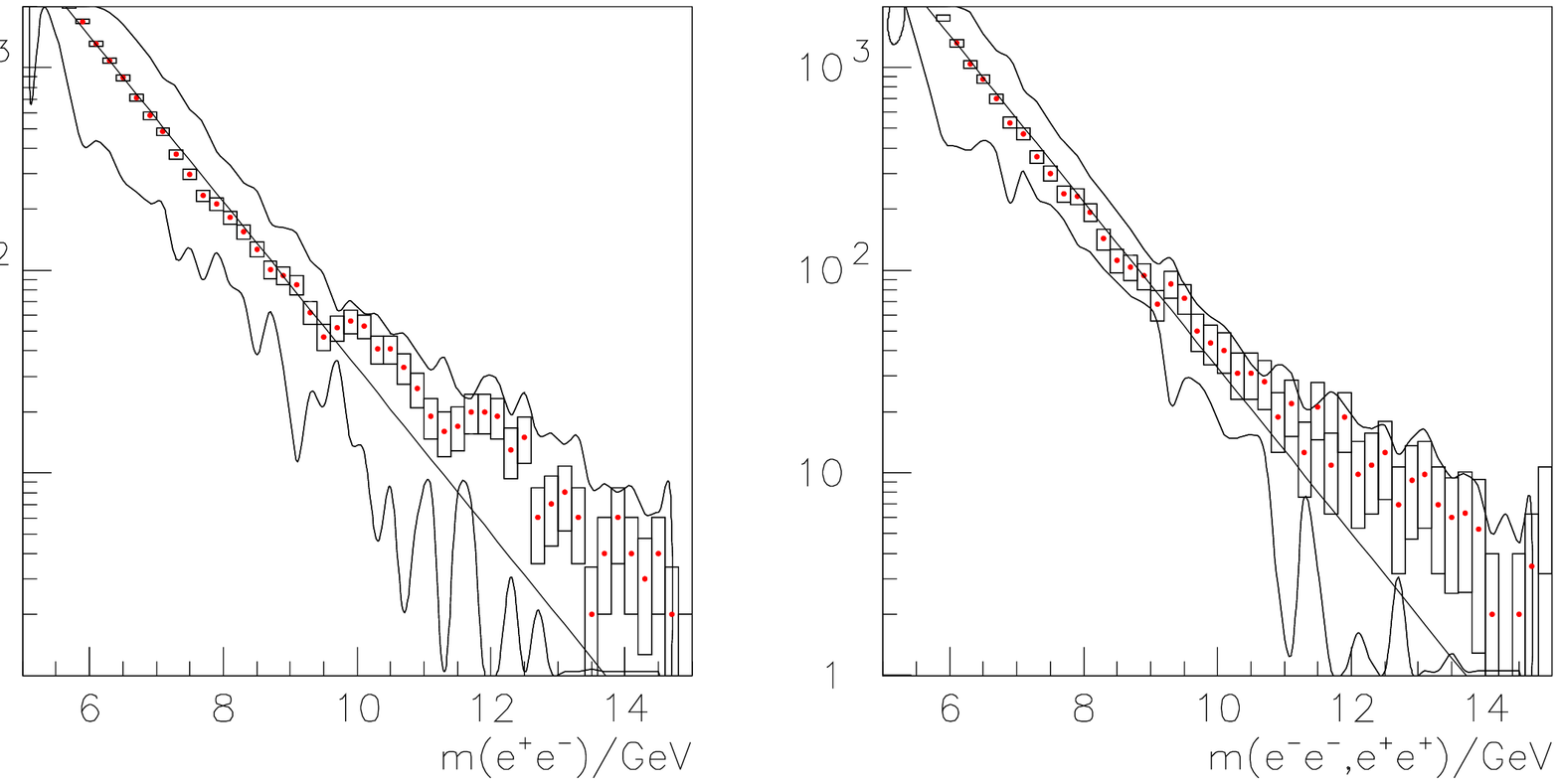}
\insertplot{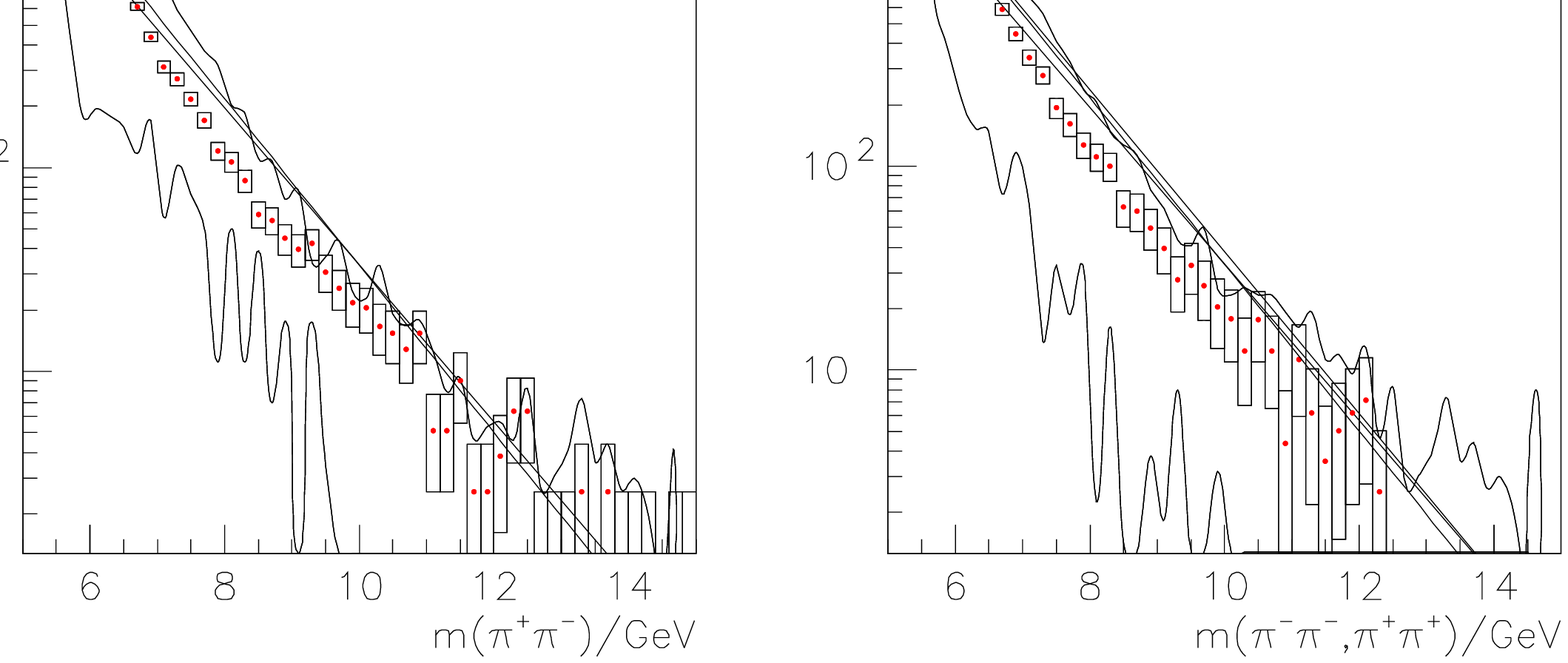}
\insertplot{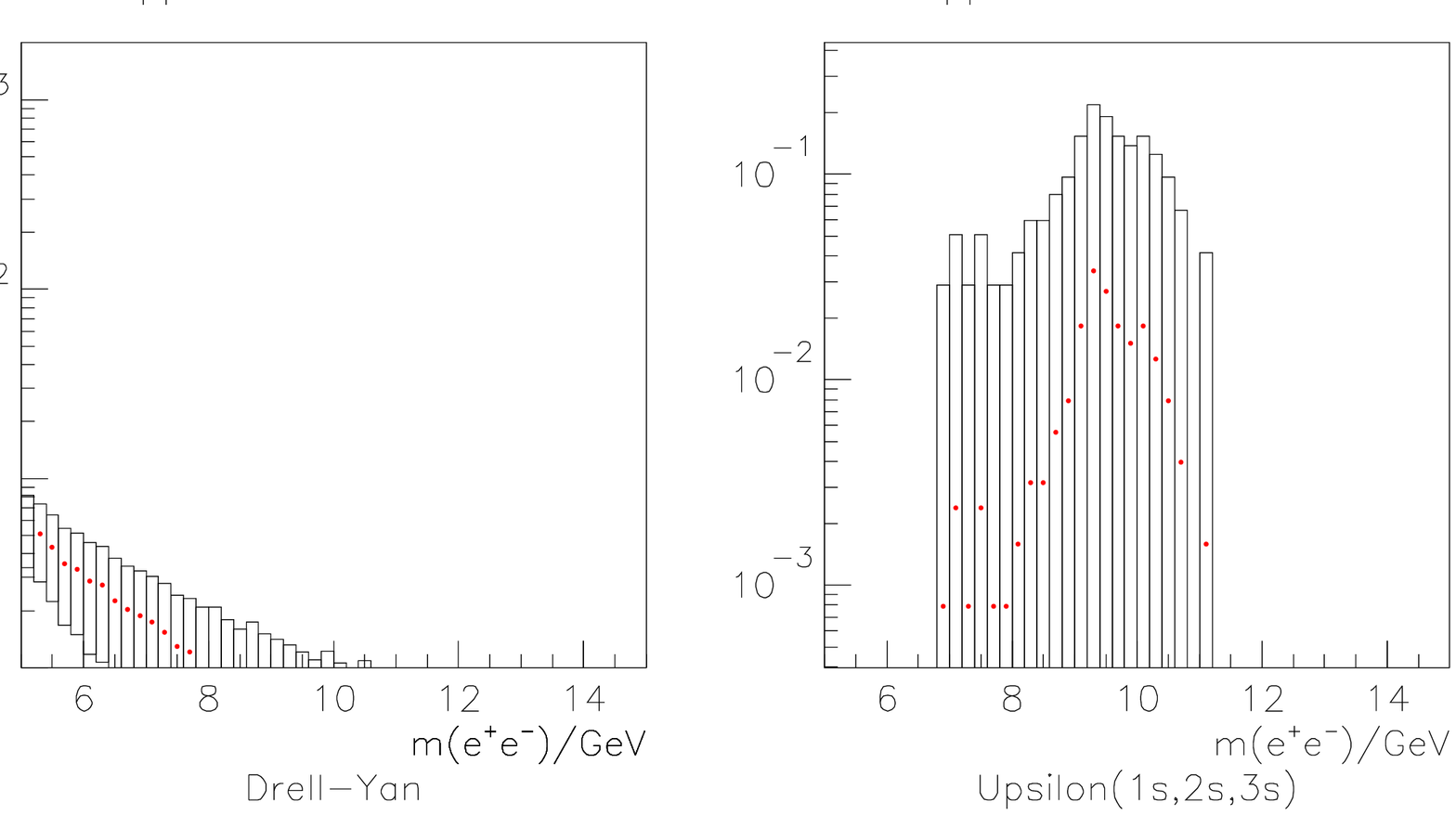}
\vspace*{-6.5cm}
\caption[]{Invariant mass for di-electrons and di-pions (control sample),
for experimental data and PYTHIA 6.212. Statistical errors are shown as
error bars, the white error band represents the systematic error. 
For details see Section~\ref{cresults}. Please notice different y-axis scale 
by factor 10$^{-5}$ for $\Upsilon$ {\it (bottom, right)}.}
\label{fig2}
\end{figure}

The analyzed data sample is
$N_{Event}$=2,400,060 $Au$+$Au$ 10\% most central,
additionally $N_{Event}$=212,030 $Au$+$Au$ 10\% most central, 
triggered by the Level-3 $\Upsilon$ algorithm, and 
$N_{Event}$=2,309,063 $Au$+$Au$ minimum bias events.
The data were taken during a period of $\simeq$100 days 
at $\sqrt{s}=200$~GeV.
Assuming a number of 1150 binary $p$+$p$ collisions 
per one Au+Au collision, the data set corresponds to 
$\simeq$4.2$\cdot$10$^9$ $p$+$p$ collisions.
As a comparison, the amount of raw data on tape correspond to
$\simeq$10$\times$total data sample of a recorded 
100~fb$^{-1}$ $B$ meson factory data sample.
Fig.~\ref{fig2} shows the experimental results for the reconstructed invariant mass.
A pseudorapidity cut of $|$$\eta$$|$$\leq$1 is applied.
The top row shows di-electrons
for unlike-sign {\it (left)} and like-sign\footnote{ 
The like-sign histogram is normalized according 
2$\cdot$$\sqrt{N^{++}\cdot N^{--}}$.} {\it (right)} charge pairs.
The straight line represents an exponential fit to the $m$$\leq$8~GeV region,
in order to visualize slope changes.
The second row shows invariant mass for a control sample, 
i.e.\ $\pi^{+-}$$\pi^{+-}$ candidates selected by dE/dx$\leq$15~keV/cm
and normalized to the number of di-lepton candidates.
The systematic error\footnote{All error components were summed in quadrature.} 
consists of  of a $p_T$ dependant $\Delta$$p_T$/$p_T$ resolution, 
a charged pion contamination of 17\%, 
and an additional 10\% systematic uncertainty induced by cut selection. 
The experimental data are compared to PYTHIA 6.212 \cite{pythia} calculations
for the different subprocesses
$p$$p$$\rightarrow$$B$$\overline{B}$$\rightarrow$$e^+$$\overline{\nu}$$X$$e^-$$\nu$$X$
{\it (third row, left)},
$p$$p$$\rightarrow$$D$$\overline{D}$$\rightarrow$$e^-$$\nu$$X$$e^+$$\overline{\nu}$$X$
{\it (third row, right)},
Drell-Yan $p$$p$$\rightarrow$$e^+$$e^-$$X$
{\it (fourth row, left)},
and $\Upsilon$ production 
$p$$p$$\rightarrow$$\Upsilon$(1$S$,2$S$,3$S$)$\rightarrow$$e^+$$e^-$
{\it (fourth row, right)}.
The expected STAR mass resolution is 
$\Delta$$m$=340~MeV for the $\Upsilon$(1$S$) alone, and 
$\Delta$$m$=610~MeV for the $\Upsilon$(1$S$,2$S$,3$S$) 
combined (non-resolvable).
The PYTHIA $k$-factor\footnote{
$k$-factor is defined as ratio of cross sections
for  next-to-leading-order QCD\\ \quad \quad and leading-order QCD. 
No higher loops were taken into account.} was set to $k$=3.0.
GRV94L was used as the parton distribution.
All simulated mass distributions are normalized 
to the experimental data sample (Section~\ref{cresults}),
taking into account enhancement factors on the Level-3 trigger.
Two conclusions can be drawn.

\noindent {\it (1)} The dilepton invariant mass distributions 
follow an exponential slope for $m$$\leq$9 GeV, whereas the fit 
line is in the following refered to as {\it reference line}.
An enhancement above the reference is visible for $m$$\geq$9~GeV,
for both unlike-sign and like-sign di-electrons.
The normalized di-pion control sample shows a non-exponential slope,
for $m$$\geq$6~GeV beneath the reference line.
As seen from reference line comparison,
any possible contamination from the di-pions cannot explain the observed
enhancement of the dileptons for $m$$\geq$9~GeV.
From comparison to PYTHIA, it can be concluded
that the major distribution to the di-electrons results
from $B$$\overline{B}$ production with subsequent semileptonic decays.
Drell-Yan does not contribute significantly, as the cross section is 
small $\sigma$($q$$\overline{q}$$\rightarrow$$e^+$$e^-$)$\leq$10~nb.
$D$$\overline{D}$ events are highly suppressed with a factor $\simeq$10$^7$
due to the electron $p$$\geq$3.0~GeV/c cut.

\noindent {\it (2)} The unlike-sign dilepton distribution seems to indicate 
peak-like structure in the mass region $m$$\geq$9 GeV.
In order to exclude possible accidentally used
$B$=0~T events, which could fake high-$p_T$ tracks,
a second independent analysis confirmed the peak structure.
The peaks remain visible when using a {\it 2-dim} (Level-3 trigger)
or {\it 3-dim} (offline) track finder algorithm, or when perform
track finding with or without vertex constraint.
If the peak structure was induced by $\Upsilon$ production, 
it would indicate the unlikely fact, that the production cross 
section would be enhanced by a factor $\simeq$10$^3$
compared to the baseline prediction (Section~\ref{cupsilon}).
Any additional processes generating high mass di-leptons
are not known. 


\section{Systematic studies of $\sigma_{pp}$($\Upsilon$) cross section prediction}

The question of whether the observed peak structure
might be induced by enhanced $\Upsilon$ production
was investigated by systematic variation of three different 
basic QCD parameters in PYTHIA 6.212.

\noindent {\it (1)} In the $\sigma_{pp}$($\Upsilon$) calculation 
a fixed mass $m$($b$)=4.75~GeV is assumed,
but $m$($b$) depends on the momentum transfer 
$Q^2$ in $g$$g$,$q$$\overline{q}$$\rightarrow$$b$$\overline{b}$,
i.e.\ 
$m$($b$)$\simeq$4.5~GeV for $Q$=1~GeV/c,
$m$($b$)$\simeq$3.5~GeV for $Q$=10~GeV/c,
$m$($b$)$\simeq$3.0~GeV for $Q$=100~GeV/c.

\noindent {\it (2)} $\sigma_{pp}$($\Upsilon$) also depends on the 
squared $\Upsilon$ wave function at the origin $|$$\Psi$(r=0)$|^2$,
which is calculated from the leptonic decay width

\vspace*{-0.5em}
\begin{equation}
\Gamma(\Upsilon \rightarrow l^+ l^-) = 
\frac{64\pi}{9}
\frac{\alpha_{em}^2 e^2 | \Psi (r=0) |^2}{m^2_{e^+e^-}}
\end{equation}

\noindent with exp.\ errors $\pm$4\%(1s),$\pm$16\%(2s),$\pm$18\%(3s).

\noindent {\it (3)}
$\sigma_{pp}$($\Upsilon$) depends on the strong coupling constant
$\alpha_s$, which itself depends on the gluon distribution $xG(x,Q^2(x))$.
As an example, the early stage of an $A$+$A$ collision might be governed
by gluon saturation, simply given by the geometrical condition, 
that the probability of two colliding gluons in a gluon cascade 
is equal to one.

\vspace*{-1em}
\begin{equation}
\label{glass}
\frac{3 \pi^2 \alpha_S A}{2 Q^2_{Saturation}} \times
\frac{x\cdot G(x,Q^2(x))}{\pi R_A^2} = 1
\end{equation}

\noindent Additionally, $\alpha_s$ is a function of temperature
$\alpha_s$($T$)$\sim$1/ln($T$) \cite{vogt2}. At a scale of 
$\sqrt{s}$=$\sqrt{3}$~GeV, \cite{pdg} quotes
\mbox{$\alpha_S$=0.280$\pm$0.035~(expt)$\pm$~0.050~(syst)$^{+0.035}_{-0.030}$~(theory)}.
The total (quadratically summed) error is $\pm$0.069, thus 
$\Delta$$\alpha_S$/$\alpha_S$=25\%.

\noindent The variation of 
$\Delta$$\sigma_{pp}$($\Upsilon)$/$\sigma_{pp}$($\Upsilon)$
from the baseline prediction (Ch.~\ref{cupsilon})
was investigated using PYTHIA 6.212, and found to be
a factor 1.0$-$3.0 (only an increasing effect observed) for $m$($b$),
a factor 0.7$-$1.3 for $|$$\Psi$(r=0)$|^2$, and
a surprisingly large factor of 10$^{-2}$$-$1.0 (only a decreasing effect observed)
for $\alpha_S$.
However, in order to explain the observed peak structure in Fig.~\ref{fig2} (top, left)
a $\sigma_{pp}$($\Upsilon)$-enhancement of a factor $\simeq$10$^3$ would be needed. 
Thus, at this stage, $\Upsilon$ production must be excluded as explanation. 
For alternative explanations, 
two additional effects should be considered:

\noindent {\it (1)} A peak structure might be generated by $B$$\overline{B}$ ($m$=10.56~GeV)
or $B_s$$\overline{B_s}$ ($m$=10.74\\ 
GeV) threshold effects. Both meson species
decay with large branching fractions of 10.5\% and 8.1\%, respectively.
However, due to the undetected $\nu$ in each decay 
and additionally hard-to-calculate QCD threshold dynamics,
the size of the effect is difficult to estimate.

\noindent {\it (2)} A $B$ meson oscillates by second-order weak interaction 
into a $\overline{B}$ according to 
cos($\Delta$$m$$t$), with $\Delta$$m$=0.503~ps$^{-1}$ \cite{nick}.
Using a $B$ meson lifetime of $c$$\tau$=462~$\mu$m \cite{pdg} 
and a typical $B$ meson Lorentz boost  
of $\gamma$=1.2 (corresponds to $E_{kin}$($B$)=1~GeV)
it can be estimated for STAR, that $\simeq$2/3 of all $B$ remain a $B$, 
and $\simeq$1/3 oscillate into a $\overline{B}$.
Every oscillation flips the sign of the lepton charge 
in the semileptonic decay, thus the oscillation could be observed
in an unlike-sign/like-sign charge asymmetry vs.\ $B$ meson decay vertex.
Unfortunately, the STAR $z$ vertex resolution has not been sufficient 
so far to resolve the oscillation.

\vspace*{-0.5em}
\section{Apex}

Explanations of the observed enhancement and peak structures
in the electron invariant mass distributions might only be possible 
with a largely increased $Au$+$Au$ data sample, which will be collected in 2004.
Lepton identification at STAR will be improved by an 
electromagnetic calorimeter with $|$$\eta$$|$$\leq$2 and 
$\Delta$$\varphi$=2$\pi$ coverage.
For the envisaged RHIC-2 upgrade, starting possibly from 2007,
a luminosity increase for $Au$+$Au$ by a factor 40 
to ${\cal L}$=8$\times$10$^{27}$~cm$^{-2}$$s^{-1}$ is planned,
thus the $\Upsilon$ production will not be a rare process
anymore.
Additionally a new 2-layer pixel detector is planned at STAR, 
using a total of 9$\times$10$^7$ pixels of 20~$\mu$m$^2$ in area.
Such a device will clearly enable 
an excellent measurement of $B$ meson secondary vertices.

\vfill\eject


\vspace*{-0.5em}
\begin{thebibliography}{99}  
\bibitem{star} STAR Collaboration, Nucl.~Phys.~{\bf A661}(1999)681
\bibitem{early} B.\ Zhang et al., Phys.~Rev.~{\bf C65}(2002)054909, nucl-th/0201038
\bibitem{kolb} U.\ Heinz, P.\ Kolb, Nucl.~Phys.~{\bf A702}(2002)269, hep-ph/0111075 
\bibitem{vogt1} R. Vogt, J.~Phys.~{\bf G23}(1997)1989
\bibitem{pdg} Particle Data Group, Phys.~Rev.~{\bf D66}(2002)010001
\bibitem{cdf} CDF Collaboration, Phys.~Rev.~Lett.~{\bf 88}(2002)161802
\bibitem{cleo} CLEO Collaboration, hep-ex/0207060
\bibitem{marek} M.~Ga\'zdzicki, M.~I.~Gorenstein, Phys.\ Lett.\ {\bf B517}(2001)250)
\bibitem{vogt2} J.~F.~Gunion, R.~Vogt, Nucl.~Phys.~{\bf B492}(1997)301, hep-ph/9610420
\bibitem{emc} T.~Cormier et al., Nucl.~Inst.~Meth.~{\bf A483}(2002)734, hep-ex/0107081
\bibitem{l3} J.~S.~Lange et al., Nucl.~Inst.~Meth.~{\bf A453}(2000)397  
\bibitem{pythia} T.~Sj\"ostrand et al., Comp.~Phys.~Comm.~{\bf 153}(2001)238, hep-ph/0010017 
\bibitem{nick} BELLE Collaboration, Phys.~Rev.~{\bf D 67}(2003)052004, hep-ex/0212033
\end{thebibliography}
\end{document}